\documentclass[pra,twocolumn,superscriptaddress,showpacs,amsmath,amstex,amssymb,citeautoscript]{revtex4-1}

\usepackage[colorlinks=true,citecolor=blue,linkcolor=blue,breaklinks=true]{hyperref}

\usepackage{color,graphicx}
\usepackage{bm}
\usepackage{amsmath}
\usepackage{amssymb}
\usepackage{mathrsfs}
\usepackage{times}
\allowdisplaybreaks
 \usepackage{caption}
\usepackage{subcaption}

\usepackage[T1]{fontenc}
\usepackage[utf8]{inputenc}
\usepackage{lipsum}
\usepackage{amsmath}
\usepackage{amssymb}
\usepackage{bbm}
\usepackage{braket}
\usepackage{xcolor}
\usepackage{pifont}
\usepackage[mathscr]{euscript}
\usepackage[shortlabels]{enumitem}
\usepackage[justification=justified]{subcaption}
\usepackage{graphicx}
\usepackage{lipsum}
\allowdisplaybreaks
\usepackage{float}
\usepackage{graphicx}
\usepackage{dsfont}
\usepackage{comment}
\usepackage[colorlinks=true]{hyperref}  
\hypersetup{
    bookmarks=true,         
    unicode=false,          
    pdftoolbar=true,        
    pdfmenubar=true,        
    pdffitwindow=false,     
    pdfstartview={FitH},    
    pdftitle={Anderson localization in doped semiconductors},    
    pdfauthor={Poduval and Das Sarma},     
    pdfsubject={},   
    pdfcreator={},   
    pdfproducer={}, 
    pdfkeywords={} {} {}, 
    pdfnewwindow=true,      
    colorlinks=true,       
    linkcolor=blue, 
    citecolor=blue,        
    filecolor=magenta,      
    urlcolor=blue           
}

\newcommand{\bk}{\vec{k}}
\newcommand{\bq}{\vec{q}}

\newcommand{\equref}[1]{Eq.~(\ref{#1})}

\newcommand{\figref}[1]{Fig.~\ref{#1}}

\renewcommand{\approx}{\simeq}

\renewcommand{\vec}[1]{\boldsymbol{#1}}

\definecolor{wrongultramarine}{rgb}{1,0.5,0}

\begin{document}
	
\newcommand {\beq} {\begin{equation}}
\newcommand {\eeq} {\end{equation}}
\newcommand {\bqa} {\begin{eqnarray}}
\newcommand {\eqa} {\end{eqnarray}}
\newcommand {\ca} {\ensuremath{c^\dagger}}
\newcommand {\Ma} {\ensuremath{M^\dagger}}
\newcommand {\psia} {\ensuremath{\psi^\dagger}}
\newcommand {\fbar} {\ensuremath{\bar{f}}}
\newcommand {\psita} {\ensuremath{\tilde{\psi}^\dagger}}
\newcommand{\lp} {\ensuremath{{\lambda '}}}
\newcommand{\A} {\ensuremath{{\bf A}}}
\newcommand{\QQ} {\ensuremath{{\bf Q}}}
\newcommand{\kk} {\ensuremath{{\bf k}}}
\newcommand{\qq} {\ensuremath{{\bf q}}}
\newcommand{\kp} {\ensuremath{{\bf k'}}}
\newcommand{\rr} {\ensuremath{{\bf r}}}
\newcommand{\rp} {\ensuremath{{\bf r'}}}
\newcommand {\ep} {\ensuremath{\epsilon}}
\newcommand{\nbr} {\ensuremath{\langle r r' \rangle}}
\newcommand {\no} {\nonumber}
\newcommand{\up} {\ensuremath{\uparrow}}
\newcommand{\dn} {\ensuremath{\downarrow}}
\newcommand{\rcol} {\textcolor{red}}
\newcommand{\bcol} {\textcolor{blue}}
\newcommand{\lt} {\left}
\newcommand{\rt} {\right}


\title{Anderson localization in doped semiconductors }
\author{Prathyush P. Poduval}
\affiliation{Condensed Matter Theory Center and Joint Quantum Institute, Department of Physics, University of Maryland, College Park, MD 20742, USA}

\author{Sankar Das Sarma}
\affiliation{Condensed Matter Theory Center and Joint Quantum Institute, Department of Physics, University of Maryland, College Park, MD 20742, USA}

\date{\today }

\begin{abstract}
    We theoretically consider the problem of doping induced insulator to metal transition in bulk semiconductors by obtaining the transition density as a function of compensation, assuming that the transition is an Anderson localization transition controlled by the Ioffe-Regel-Mott (IRM) criterion. We calculate the mean free path, on the highly doped metallic side, arising from carrier scattering by the ionized dopants, which we model as quenched random charged impurities. The Coulomb disorder of the charged dopants is screened by the carriers themselves, leading to an integral equation for localization, defined by the density-dependent mean free path being equal to the inverse of the Fermi wave number, as dictated by the IRM criterion. Solving this integral equation approximately analytically and exactly numerically, we provide detailed results for the localization critical density for the doping induced metal-insulator transition. 
\end{abstract}

\pacs{}

\maketitle

\section{Introduction}
Semiconductors, (e.g. Si, Ge, GaAs, InAs, InSb and others) are small band gap insulators which can be doped by suitable dopant atoms to produce metallic free carriers in the conduction or valence band, depending respectively on whether the dopants are donors or acceptors \cite{ashcroft2022solid,shockley1950,smith1978,mott1990,shklovskii2013electronic}. In the current work, we discuss the metallicity of (donor-) doped semiconductors in terms of conduction band electrons as the carriers in a generic sense, but our work should apply equally well to valence band hole doping by acceptors. At finite (e.g. room) temperatures, electrons are thermally excited from the dopants to the conduction band to act as free carriers, and the rather low conduction band carrier density ($\sim 10^{18}-10^{21} \rm cm^{-3}$) enables controlling their electrical properties with external gate voltages leading to the modern microelectronics industry based on semiconductor (mostly Si) transistors. Our work focuses on low temperatures, specifically $T=0$, where increased doping could lead to a doping-induced insulator (at low or no doping) to metal (at high doping) transition, which has been extensively studied both experimentally and theoretically over the years\cite{mott1990,shklovskii2013electronic}. Thermal carrier excitations from the dopants do not play any role in such a $T=0$ metal-insulator transition (MIT), with the metal (insulator) occurring at high (low) density, and this doping induced MIT is thought to be a quantum phase transition of paradigmatic importance in condensed matter physics, which has been studied a great deal over the years\cite{mott1968metal,lee1985disordered,imada1998metal,alexander1968semiconductor,yamanouchi1967electric}.

In spite of extensive research, the key issue of the fundamental nature of the actual experimental low temperature MIT in doped semiconductors is not settled. There are (at least) three possible mechanisms which could lead to the doping induced MIT in semiconductors: interaction-induced Mott transition in the impurity band, Coulomb disorder induced percolation transition, and random disorder induced Anderson localization. (There are in fact other theoretical possibilities such as structural transitions leading to MIT\cite{furubayashi1994structural,mineshige1996crystal,mcwhan1969mott,jullien1971etude} among others\cite{di2017disorder}, which we ignore because they are unlikely to be operational in simple doped semiconductors of interest in our work.) In the Mott transition scenario, the preferred scenario in most modern discussions on the doping induced MIT, the dopant electrons form an impurity band close to the conduction band, and increasing doping density leads to an electron-electron interaction induced insulator-to-metal transition as envisioned by Mott long time ago\cite{mott1949basis,mott1990} as the Coulomb interaction between the electrons gets screened out. The critical density for the Mott transition $n_M$ is universally accepted to be roughly given by \cite{shklovskii2013electronic,chen2016metal,edwards1978universality}:
\beq
n_M\sim 0.02/a_B^3,
\label{eq:mott}
\eeq
where $a_B$ is the effective Bohr radius for the host semiconductor. The percolation transition, which is semi-classical (and essentially a carrier `trapping' mechanism by the disorder potential fluctuations, and is in fact a classical version of Anderson localization in smooth long-range disorder), arises from the smooth background disorder potential due to the random ionized dopants leading to an inhomogeneous `mountain and lake' landscape consisting of spatial puddles of conducting electrons within a disordered insulating background. In such a scenario, electron transport occurs through percolating conducting paths between different puddles in the inhomogeneous system. As a result, the electrons can conduct through the bulk only if the Fermi level is high enough for a percolation path to exist through the whole sample in the inhomogeneous disorder landscape\cite{shklovskii2013electronic,kirkpatrick1973percolation}. The actual percolation critical density is non-universal, being dependent on many details of the system, and can only be numerically calculated approximately \cite{efros1976critical,sarma2011electronic,arnold1974disorder,sarma2013two}. Many doped two-dimensional semiconductor systems have been shown to manifest the percolation MIT because of the dominance of Coulomb disorder\cite{leturcq2003resistance,ilani2000unexpected,lilly2003resistivity,leturcq2003resistance,manfra2007transport,tracy2009observation,jiang1988threshold}. For 3D doped semiconductors, an approximate estimate for the percolation critical MIT density $n_p$ is given by \cite{shklovskii2013electronic,spinelli2010electronic,huang2021metal}:
\beq
n_p\sim 0.7 n_i^{2/3} /a_B,
\label{eq:perc}
\eeq
where $n_i$ is the random charged impurity density (i.e. ionized dopant density). {This expression for $n_p$ is valid in the strongly doped regime, with $n_i a_B^3\gg1$.} Finally, the Anderson localization transition is purely quantum, and arises from the destructive interference of electron waves induced by the disorder potential created by the random quenched ionized dopant impurities, provided that the random disorder is strong enough \cite{anderson1958absence,shklovskii2013electronic,mott1990}. The Anderson localization critical density for doped semiconductors is neither universal nor easy to calculate accurately as it depends on the microscopic details of the random disorder (which would vary from sample to sample). But a reasonable estimate for the localization sets in when the electron loses coherence because of momentum scattering, leading to the disorder induced broadening of the electron momentum itself {{becoming}} equal to the Fermi momentum. Thus, $n_c$ for localization is defined implicitly by the following equation:
\beq
k_F\cdot l_{MFP} =1. 
\label{eq:irm}
\eeq
Since both $k_F$ and $l_{MFP}$ depend on the carrier density $n$, \equref{eq:irm} provides $n_c$ directly if $l_{MFP}$ is known.  Eq. \ref{eq:irm} defines the well-established IRM criterion for Anderson localization, where the loss of coherence due to disorder scattering is identified as the localization transition from a metal to an insulator.

There is considerable debate on whether the MIT in a particular sample proceeds through a Mott or an Anderson mechanism. In many uncompensated semiconductors, the critical density apparently agrees with the condition for Mott transition [\equref{eq:mott}], signifying that electron-electron correlations presumably play an important role in the MIT\cite{edwards1978universality,alexander1968semiconductor,zylbersztejn1975metal,mcwhan1970metal,kachi1973metal}. On the other hand, theoretical and optical studies of the same systems often show that the transition may be of Anderson type, since the impurity band is separated from the conduction band, and there is no Hubbard gap formation\cite{gaymann1995temperature,bhatt1981single}. This implies that in most systems the MIT is caused by a combination of Mott and Anderson mechanisms, making the theoretical problem complicated as disorder and correlation are most likely equally important. One key problem is that the disorder usually cannot be controlled effectively, and significant amount of disorder is required to overcome the Coulomb energy responsible for the Mott transition mechanism. However, recently, phase-change materials (PCM) like GeSb$_2$Te$_4$ and  Li$_x$Fe$_7$Se$_8$ have been studied extensively, where the MIT is shown to be entirely caused by Anderson localization\cite{siegrist2011disorder,ying2016anderson,bragaglia2016metal,wang2020resistive,rostami2020review,shekhawat2022improved,shekhawat2022improved}. The disorder in PCM systems can be controlled by varying the annealing temperature, enabling the tuning of the system across the MIT at high disorder. Additionally, an Anderson type MIT was also reported in SrNbO$_{3-x}$N$_x$, where the disorder is controlled by the added Nitrogen in the system\cite{oka2021electron}. The study of the PCM materials\cite{siegrist2011disorder,ying2016anderson,bragaglia2016metal,wang2020resistive,rostami2020review,shekhawat2022improved,shekhawat2022improved} unambiguously establishes the principle that Anderson localization by itself could drive the doping-induced MIT in semiconductors without Mott correlation effects playing a major role.  In addition, carrier density-tuned MIT in 2D semiconductors has been interpreted as an Anderson localization induced crossover phenomenon\cite{sarma2014two}.

In the current work, we make the uncritical assumption of the doped semiconductor MIT to be arising from the Anderson localization at the bottom of the semiconductor conduction band (or in the impurity band), and obtain $n_c$ as a function of compensation $K$, which assumes that only a fraction  $(1-K)$ of the donor (acceptor) atoms are actually ionized with the rest being compensated by the presence of random acceptors (donors) in the environment. Compensation introduces considerable random disorder into doped semiconductors, facilitating the Anderson localization driven MIT. Sometimes the compensation is deliberately introduced in the system by doping the sample simultaneously with both donors and acceptors, but often compensation happens unknowingly simply because the sample would typically have both donors and acceptors. Note that for $K=1$, the sample is fully compensated, and in principle, there are no free carriers, leading to a non-conducting insulating phase independent of the doping level. 

Our model for Anderson localization is non-interacting in a direct sense as the problem of localization in the presence of both interaction and disorder is intractable, but screening of the charged disorder by the electrons themselves is included non-perturbatively in the theory through the static random phase approximation (RPA) using the finite momentum Lindhard function\cite{lindhard1954kgl,lindhard1954properties}. Thus, the effects of interaction are included indirectly in the theory through the screening mechanism of disorder, which arises strictly from the mutual electron-electron Coulomb interaction. The physical picture addressed in our theory is simple: We calculate the carrier mean free path (assuming metallic electrons in the high-doping limit) due to the screened Coulomb disorder scattering by random charged impurities, and then equate the mean free path to the inverse Fermi momentum to obtain the critical MIT density $n_c$. Our theory is valid only in the metallic regime for carrier density $n>n_c$, but we extend it all the way to $n=n_c$ (from above) in order to estimate the critical density. The theory is obviously approximate as we use the Boltzmann transport theory assuming Born approximation and employ the Lindhard-RPA screening in order to obtain the effective Coulomb disorder. We emphasize that the inclusion of screening is essential in the theory since unscreened Coulomb disorder gives a logarithmically divergent scattering rate in three dimensions and screening regularizes the singular bare scattering in a parameter free manner. 

One sharp experimentally testable difference between our Anderson localization scenario and the Mott transition scenario is that the Mott transition implies a universal critical density defined by \equref{eq:mott} which is determined entirely by the host semiconductor effective Bohr radius whereas the localization critical density is non-universal, and should show sample to sample variations depending on the (often unknown) level of intrinsic compensation. Experimentally, the MIT critical density is sample dependent, arguing against a pure Mott transition picture for the doping induced MIT. This is understandable as the Mott transition considers an unphysical perfectly ordered lattice arrangement of the dopants assuming the full ionization of all dopants. In reality, the ionization is unlikely to be complete and certainly there is considerable spatial randomness in the dopant locations, both aspects of physics ignored in the Mott transition scenario. Our work focuses entirely on the localization scenario as defined by the IRM criterion in \equref{eq:irm}, and we obtain the critical transition density defined by \equref{eq:irm} as a function of system parameters. 

The rest of this article is organized as follows. In Section~\ref{sec:model}, we provide the basic theory and the model we use. Then, in Section~\ref{sec:numericalnc}, we provide approximate analytical results for the relationship between $n_c$ and the impurity density $n_i$ in the high and low density limit, along with exact numerical results, to understand the dependence of $n_c$ on various model parameters. In Section~\ref{sec:analytic}, we provide the corresponding analytical results for $n_c$ as a function of compensation $K$, alongside a comparison with numerical results. 
We then compare the critical density from Anderson localization with the critical density from Mott criterion and the percolation criterion, to predict the range of compensations where we expect the MIT to be primarily of Anderson type. We conclude in Section~\ref{sec:conclusion} with a summary and discussion of the open questions. 

\section{ Model and theory}
\label{sec:model}
We consider a 3D electron (or hole) system at $T=0$ modeled by a parabolic band structure $\epsilon_{\bk}=\hbar^2 k^2/2m$, where $\bk$ is the 3D wavevector. The system is characterized by the effective mass $m$, the 3D carrier density $n$, the background dielectric constant $\kappa$ for the screened 3D Coulomb interaction, and the average density of the random charged impurity $n_i$. In addition to these parameters, we also have the degeneracy of the band $g=g_sg_v$, where $g_s$ is the spin degeneracy and $g_v$ is the valley degeneracy. We will show all the analytic results for a generic degeneracy, however, our numerical results will be for $g_s=2$ and $g_v=1$ unless mentioned otherwise. We will also use $m=0.4m_e$ and $\kappa=12$ as the typical system parameters for numerical results (with Si being the appropriate semiconductor).

The Coulomb interaction in 3D is given by $V_{\bq} = \frac{4\pi e^2}{\kappa q^2}$, and the resulting screened electron-impurity interaction is given by 
\beq
u_{\bq}=\frac{V_{\bq}}{\epsilon(q)} = \frac{4\pi e^2}{\epsilon(q)\kappa q^2},
\eeq
where $\epsilon(q)$ is the static RPA screening function in 3D which is given by 
\beq
\epsilon(q) =1+\frac{4\pi e^2}{\kappa q^2}\Pi(q),
\eeq
where $\Pi(q)$ is the 3D static polarizability function given by\cite{coleman2015introduction}
\beq
\Pi(q) = \frac{g mk_F}{2\pi^2\hbar^2}F(q/2k_F),
\eeq
where $F(x)$ is 
\beq
F(x)=\frac{1}{4x}\lt( \lt( 1-x^2\rt) \ln\lt|\frac{x+1}{x-1}\rt| \rt) +\frac12.
\eeq
Combining these expressions, the static RPA screening function is written as 
\beq
\epsilon(q)=1+\lt(\frac{q_{TF}}{q}\rt)^2F(q/2k_F),
\eeq
where $q_{TF}$ is the $3D$ Thomas-Fermi vector given by 
\beq
q_{TF} = \sqrt{\frac{2gmk_Fe^2}{\pi\kappa\hbar^2}}.
\eeq

The key variable to estimate in the IRM criterion is the mean free path $l_{MFP}$. The mean free path $l_{MFP}$ is defined by the Fermi velocity and a characteristic scattering time scale $\tau$ as $l_{MFP}=v_F\tau$. Traditionally, $\tau$ is defined to be the transport scattering time of the carriers. However, recently, an alternative choice for $\tau$ as the quantum scattering time has been proposed\cite{ahn2022anderson}. The quantum scattering time is the momentum lifetime of the carrier, defined via the imaginary part of the self energy. Since it relates to the coherence of the carriers themselves, it is a natural choice to define the scattering time. For low carrier density, strong screening limit, using either the quantum or transport scattering time gives consistent results, however, they differ in the high density limit due to the vertex corrections that enter the transport scattering time. We will show our results using both the quantum and transport scattering time, and discuss how they affect the corresponding Anderson localization critical density.

The transport scattering time $\tau_t$ is given by  
\beq
\frac{1}{\tau_t} =\frac{2\pi n_i}{\hbar} \int \frac{d^3\bk}{(2\pi)^3}|u_{\bq}|^2(1-\cos \theta_{k,k_F})\delta(\epsilon_{\bk} - \epsilon_F),\label{eq:taut}
\eeq
where $\theta_{k,k_F}$ is the angle between $\bk$ and $\bk_F$ and $\bq=\bk-\bk_F$. We call the corresponding IRM criterion as the transport IRM criterion. Alternatively, we also consider $\tau$ to be the single particle quantum scattering time $\tau_q$, given by  
\beq
\frac{1}{\tau_q} =\frac{2\pi n_i}{\hbar} \int \frac{d^3\bk}{(2\pi)^3}|u_{\bq}|^2\delta(\epsilon_{\bk} - \epsilon_F).\label{eq:tauq}
\eeq
We call the corresponding IRM criterion as the quantum IRM criterion, which uses the quantum scattering time to define the mean free path.

Note that Eqs.~\ref{eq:taut} and \ref{eq:tauq} differ by the vertex correction factor ($1-\cos\theta$), indicating that forward scattering is suppressed for the transport scattering rate. The expression for the two quantities, $\tau_t$ and $\tau_q$, differ by the $1-\cos\theta$ term in $\tau_t$, which arises from the vertex corrections in transport quantities. The $1-\cos\theta$ term takes into account backward scattering with a higher weight as compared to forward scattering, however, for isotropic $s-$wave scattering processes the transport time and quantum time are equal, since the vertex correction for $s-$wave scattering becomes just unity. The simplified expression for the scattering times is given by, following some straightforward algebra,
\beq
\frac{1}{\tau_{t,q}} = \frac{n_im}{8\pi \hbar^3k_F^3}\left(\frac{4\pi e^2}{\kappa}\right)^2\int_0^1\frac{\gamma_{t,q} x^{2\gamma_{t,q}-1}}{\left(x^2+q_s^2F(x)\right)^2} dx,\label{eq:tausimple}
\eeq
where $\gamma_t=2$, $\gamma_q=1$ and,
\beq
q_s = \frac{q_{TF}}{2k_F}=e^\delta n^{-\frac16}, \phantom{111} \delta = \ln\left(\sqrt{\frac{gme^2}{2\pi\kappa\hbar^2}}\left(\frac{g}{6\pi^2}\right)^\frac16\right).
\eeq
In the limit of strong screening ($q_s\gg 1$), $\tau_{t}$ and $\tau_q$ are \textit{approximately} equal, since the strongly screened Coulomb interaction is approximately $s-$wave. However, the 3D RPA screening function is in reality modulated by the Friedel oscillation form factor $F(x)$, where $x=2k_F\sin\frac{\theta}{2}$ and $\theta$ is the scattering angle\cite{sarma1985single}. The value of $F(x)$ ranges from $F(x)=1$ at $x=0$ to $F(x)=0.5$ at $x=1$. Since the scattering integral is inversely related to $F(x)^2$, back scattering at large $x\sim 1$ ($\theta\sim \pi$) is enhanced over forward scattering at small $x\sim 0$ ($\theta\sim 0$). Due to this enhancement, and since $1/\tau_t$ takes into account back scattering with a higher weight than forward scattering, $\tau_q$ is larger than $\tau_t$ in the low density, high screening limit. Numerically, we find $\tau_q/\tau_t\approx 1.2$, and this ratio remains constant when $q_s \gg 1$. We note that in 2D semiconductors, on the other hand, $\tau_q/\tau_t=1$ holds exactly for strong screening at $T=0$, because the 2D RPA screening function is exactly $s-$wave for scattering that occurs on the Fermi surface\cite{ahn2022anderson,sarma1985single,sarma2014mobility}. In the opposite limit of $q_s\ll 1$, the ratio between $\tau_t$ and $\tau_q$ for 3D semiconductors is approximately calculated by keeping $F(x)=1$ in \equref{eq:tausimple}, to get,
\begin{align}
    \frac{\tau_q}{\tau_t} = \frac{\int_0^1 \frac{2x^3}{(x^2+q_s^2)^2}dx}{\int_0^1\frac{x}{(x^2+q_s^2)^2}dx} \approx -{4q_s^2\ln q_s}.
\end{align}
Considering band parameters for Si with $g_s=2,g_v=1,m=0.4m_e,\kappa=12$, the value of $\tau_q/\tau_t$ for $n\sim 10^{18}-10^{21}{\rm cm^{-3}}$ is correspondingly $\tau_q/\tau_t\sim  0.1-0.005$. Thus, in general, $\tau_t\gg \tau_q$ except in the extreme strong screening limit of $q_{TF}\gg k_F$ where $\tau_q \sim 1.2 \tau_t$.

Having established our model of screened Coulomb disorder, in the next section we proceed to provide numerical results for the critical density, and the corresponding analytical results for the critical density as a function of impurity density in the high density and low density limit, characterized by weak screening and strong screening respectively.

\section{Results for the critical density}
\label{sec:numericalnc}

In this section, we use the IRM criterion to derive the critical MIT density $n_c$ as a function of the random charged impurity density $n_i$. The corresponding equation to solve is an integral equation, and can only be solved numerically for a general density. Nevertheless, we do provide analytical results for $n_c$ as a function of $n_i$ in various limits, and study the dependence of $n_c$ on the model semiconductor parameters $g, \kappa, $ and $m$. 

Using the expression for $\tau_{q,t}$, we find the critical density $n_c$ for a given $n_i$ using the IRM criterion, which is re-stated in terms of the scattering transport time as 
\beq
\tau_{t,q}\epsilon_F=\hbar,
\eeq
where $\epsilon_F=\hbar^2k_F^2/2m$ is the Fermi energy. We call the IRM criterion that uses $\tau_t$ and $\tau_q$ as the transport and quantum IRM respectively. Additionally, we also calculate the exponent $\alpha$ defined by $\alpha = \partial\ln n_c/\partial\ln n_i$ (such that $n_c\sim n_i^{\alpha}$). The exponent $\alpha$ will be a precise theoretical prediction that can be compared with experimental results. Moreover, $\alpha$ will be different for the $n_c$ calculated via the quantum scattering time and the transport scattering time, and can be used to distinguish between the two IRM mechanisms. Qualitatively, the value of $\alpha$ is a direct measure of the strength of the screening in the system, since the screening parameter $q_s$ is related to $n_c$ as $q_s\sim n_c^{-\frac16}$, and $n_c$ increases monotonically with $n_i$. In the low density limit of $n_c\ll e^\delta$, the screening parameter $q_s\gg 1$, and thus the corresponding impurity density will be approximately equal for both the quantum IRM and the transport IRM, since the quantum scattering time and transport scattering time are approximately equal in the strong screening limit. In this limit, $\alpha$ is equal for both quantum IRM and transport IRM. In the limit of $q_s\gg 1$, we have,
\beq
    \hbar=\tau_t\epsilon_F \sim k_F^3q_s^4 \epsilon_F/n_i\sim n_c/n_i,
\eeq
and thus we get $n_c\sim n_i$, which gives $\alpha=1$ for $n_i\ll e^\delta$ ($q_s\gg 1$). In the opposite limit of $q_s\ll 1$, $\alpha$ is different for transport IRM and quantum IRM. This is because $\tau_t\sim -1/\ln q_s$, while $\tau_q\sim q_s^{2}$, when $q_s\ll 1$. For the transport IRM, we have (for $q_s\ll 1$) 
\beq
\hbar=\tau_t\epsilon_F \sim -{k_F^3 \epsilon_F /\ln q_s}{n_i}\sim n_c^{\frac53}/{n_i},
\eeq
which gives $n_c\sim n_i^\frac35$ for transport IRM in the weak screening limit. Correspondingly, for the quantum IRM, we have 
\beq
\hbar=\tau_q\epsilon_F \sim {k_F^3 \epsilon_F q_s^{2}}/{n_i}\sim n_c^{\frac43}/{n_i},
\eeq
which gives us $n_c\sim n_i^\frac34$ for quantum IRM. Thus, the dependence of the MIT density on the impurity density is weaker for the transport IRM approximation compared with the quantum IRM approximation.

\begin{figure}[tb]
    \centering
    \includegraphics[width=\columnwidth]{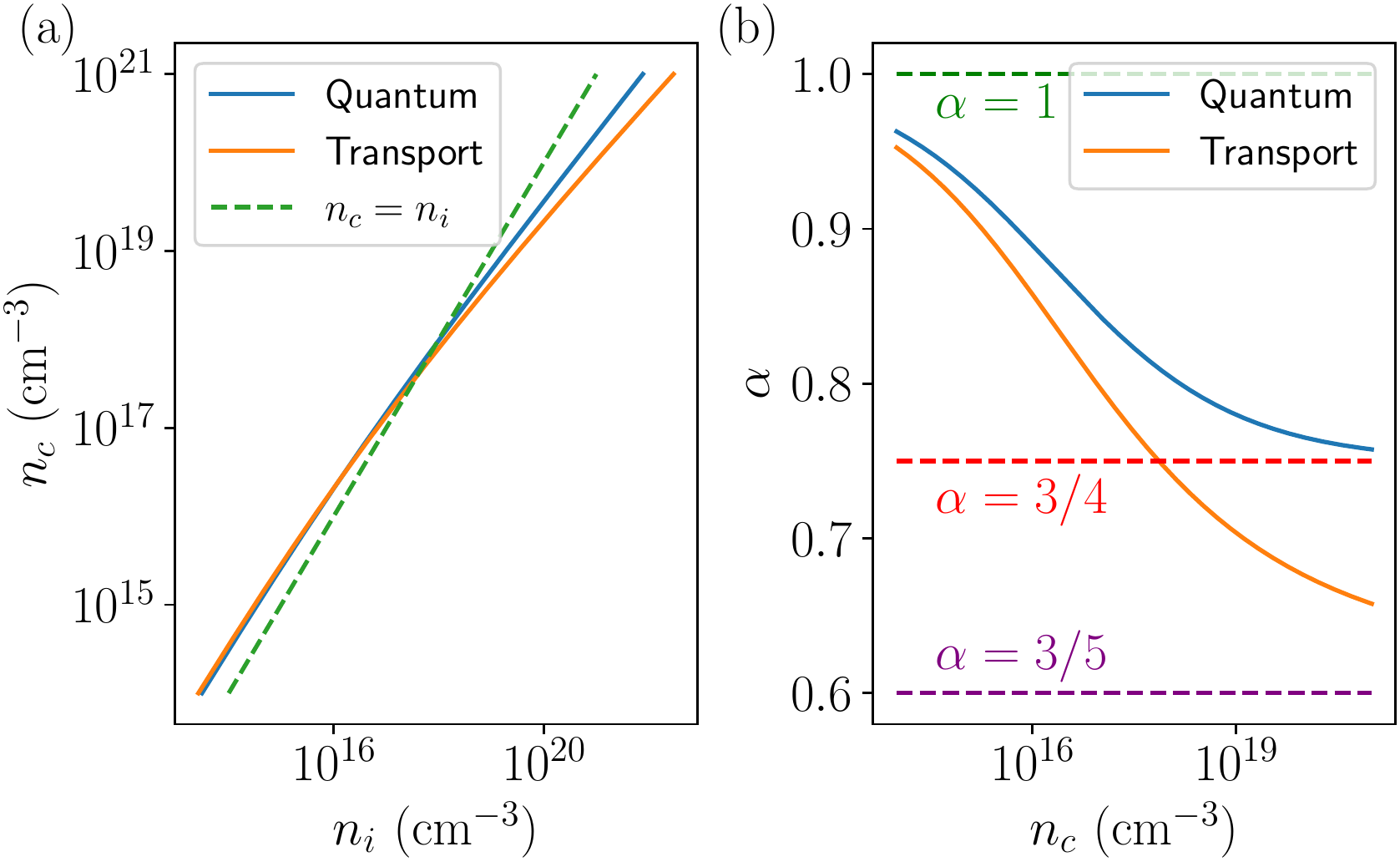}
    \caption{(a) $n_c$ as a function of $n_i$ using the quantum IRM and transport IRM criterion. There is a value of $n_i$ below which $n_c>n_i$, and is in the region of unphysical compensation. The difference between the quantum IRM and transport IRM grows as $n_i$ increases, with the $n_c$ from quantum IRM being much greater than $n_c$ from transport IRM for large $n_i$ (b) Evolution of $\alpha(n_c)$, where $\alpha = \partial\ln n_c/\partial\ln n_i$. For small $n_c$, $\alpha=1$, while for large $n_c$, $\alpha=3/4$ and $3/5$ for quantum and transport IRM respectively.}
    \label{fig:ninc}
\end{figure}

In \figref{fig:ninc}(a), we show the numerically calculated $n_c$ as a function of $n_i$ for the Si band parameters using both transport and quantum IRM. Indeed, as expected, for small $n_i$ ($q_s\gg 1$), the value of $n_c$ is linearly proportional to $n_i$ for both the transport and quantum IRM. Due to the form factor of the RPA screening function $F(x)$, the $n_c$ calculated from transport IRM is greater than the $n_c$ calculated from quantum IRM. In the other limit of large $n_i$ ($q_s\ll 1$), the $n_c$ calculated from quantum IRM is greater than the $n_c$ calculated from transport IRM. Moreover, the $n_c$ calculated from quantum and transport IRM scale with different power laws with respect to $n_i$. \figref{fig:ninc}(b) shows the corresponding evolution of $\alpha$, which monotonically decreases with increasing $n_c$. For both the quantum and transport IRM, $\alpha$ is bounded from above by $\alpha=1$. As we increase $n_i$, $\alpha$ decreases towards $\alpha=0.75$ for quantum IRM, while for transport IRM it decreases to $\alpha=0.6$ for large $n_i$. 

\begin{figure}[tb]
    \centering
    \includegraphics[width=\columnwidth]{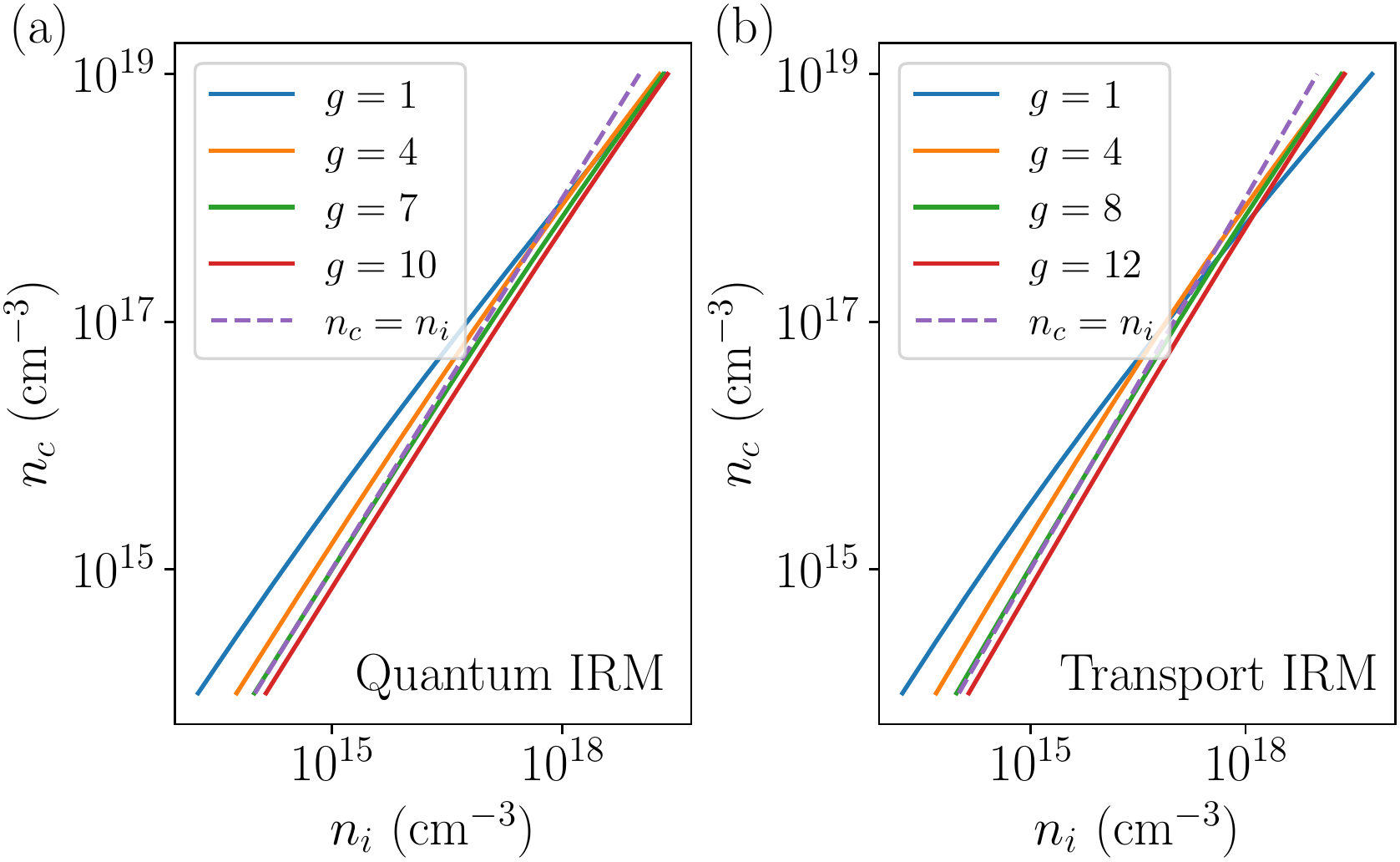}
    \caption{$n_c$ vs $n_i$ for different degeneracy $g$ calculated using (a) quantum IRM and (b) transport IRM. In the case of quantum IRM, when $g>7$, the relation $n_c<n_i$ is true for all $n_i$, while for $g<7$, there is a density $n^*$ when $n_c=n_i=n^*$, and below which $n_c>n_i$ holds. Similarly, in the case of transport IRM, when $g>8$, then the relation $n_c<n_i$ holds for all $n_i$. }
    \label{fig:nincdiffg}
\end{figure}

Unlike 2D semiconductors, the carrier density of 3D semiconductors cannot be easily independently tuned from the impurity density (since gating is typically ineffective in 3D), and since the carrier density arises from ionized dopants (which act as impurities), we have the constraint $n\le n_i$. In terms of the compensation $K$, defined via $n=(1-K)n_i$, in 3D semiconductors we always have $K\ge 0$ since, in general, there are always unintentional random donors and acceptors in the environment even when the semiconductor is deliberately doped by just one type of dopants. As we can see from \figref{fig:ninc}(a), the $K=0$ point is at the intersection of the $n_c(n_i)$ curve with the $n_c=n_i$ curve, which gives us a characteristic density $n^*$ satisfying $n_c(n^*)=n^*$. Thus, the physically valid regime is $n_i\ge n^*$. In the low density limit ($q_s\gg 1$), since the critical density goes as $n_c\sim n_i$, we can have a situation where $n_i>n_c$ is true for all $n_i$, and the Anderson localization always occurs at finite compensation. Analytically, this can be understood by evaluating the constant of proportionality between $n_c$ and $n_i$, and we find 
\beq
n_c = \frac{8\pi C_{t,q}}{3g}n_i, \phantom{11} (q_s\gg 1)\label{eq:ncnilowden}
\eeq
where 
\beq
C_{t,q}=\int_0^1\frac{\gamma_{t,q}x^{2\gamma_{t,q}-1}}{F(x)^2}dx.
\eeq
From this expression, we find that for the quantum (transport) IRM, when $g>7$ ($g>8$), the Anderson localization always occurs at finite compensation because $n_c<n_i$ is true for all $n_i$. The corresponding minimum value of the critical compensation is given by 
\beq
K_c = 1-\frac{n_c}{n_i} = 1-\frac{8\pi C_{t,q}}{3g}.
\eeq
When $g\le 7$ ($g\le 8$), the minimum compensation at which the Anderson localization occurs is $K_c=0$ for quantum (transport) IRM. In \figref{fig:nincdiffg}, we show the $n_c(n_i)$ curve for various degeneracy using the quantum IRM, and as expected we see that $n_c<n_i$ is true for all $n_i$ above a certain value of degeneracy $g$. Nominally, the conduction band in Si has $g_v=6$, and therefore, $g=12$ within the effective mass approximation, implying that any doping-induced MIT in strictly uncompensated Si conduction band (with $K=0$) cannot be an Anderson localization transition, assuming that the valley degeneracy is not lifted by random strain or other effects beyond the effective mass approximation.

\begin{figure}[tb]
    \centering
    \includegraphics[width=\columnwidth]{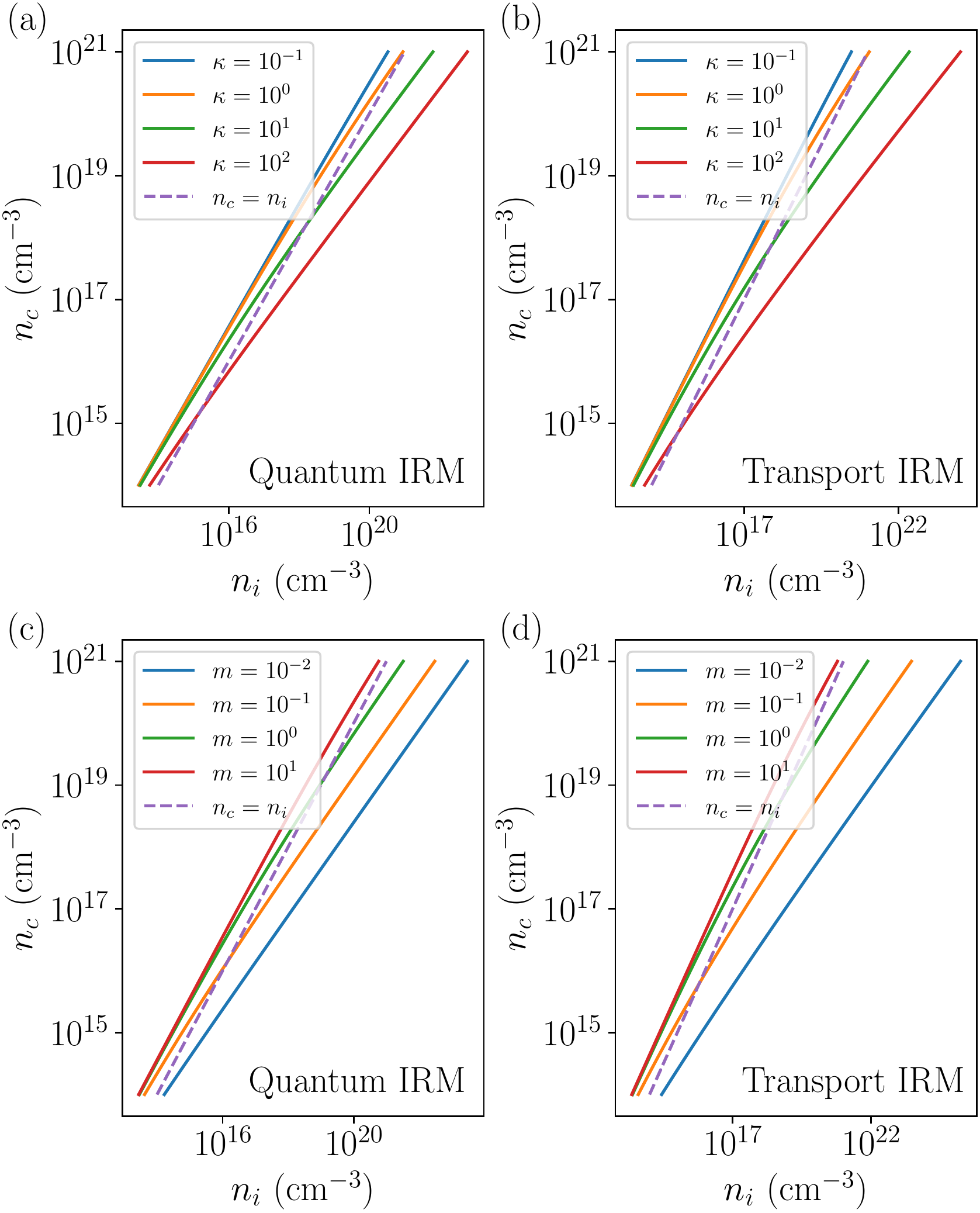}
    \caption{$n_c$ vs $n_i$ for different $\kappa$ while keeping $m=0.4m_e$ constant, calculated using (a) quantum IRM and (b) transport IRM. (c) and (d) are the same as (a) and (b), but with $m$ varied (shown in units of $m_e$) while $\kappa=12$ is kept fixed. In the case of transport IRM, the critical density goes as $n_c\sim g\lt(({m^2}/{\kappa^2}) n_i\rt)^{3/5}$, for quantum IRM it goes as $n_c\sim \lt(g^{1/3} ({m}/{\kappa}) n_c\rt)^{3/4}$. As a result, in both the quantum and transport IRM, the critical density increases with increasing $m$, and decreases with increasing $\kappa$.}
    \label{fig:nincdiffkm}
\end{figure}

For completeness, we also analyze the low density and high density dependence of $n_c$ on the effective mass $m$ and the dielectric constant $\kappa$, and the corresponding results are shown in \figref{fig:nincdiffkm}. At low density and strong screening, as we have noted before, $n_c$ is independent of $m$ and $\kappa$, and depends only on $g$ with $n_c=8\pi C_{t,q} n_i/3g$. As expected, in \figref{fig:nincdiffkm}, in the small $n_c$ limit there is little variation seen in $n_c$ as $\kappa$ and $m$ are varied.  At high density and weak screening, $n_c$ depends on $g$ and $m$ as 
\begin{align}
    n_c&\sim \left(g^\frac13 \frac{m}{\kappa}n_c\right)^\frac34,
\end{align}
for quantum IRM, while for transport IRM $n_c$ varies as
\begin{align}
    n_c&\sim g\left(\frac{m^2}{\kappa^2} n_i\right)^\frac35.
\end{align}
In the quantum IRM, $n_c$ is inversely proportional to $\kappa$, while for the transport IRM it is inversely proportional to $\kappa^{1.2}$. As expected, in \figref{fig:nincdiffkm}(a) and (b), $n_c$ does decrease with increasing $\kappa$. Similarly, in the quantum IRM, $n_c$ is proportional to $m$, while for the transport IRM it is proportional to $m^{1.2}$, and as shown in \figref{fig:nincdiffkm}(c) and (d), $n_c$ increases with increasing $m$.

We note that, by contrast, the critical MIT doping densities for the Mott and percolation transitions (defined by Eqs.~\ref{eq:mott} and \ref{eq:perc}) go respectively as $(m/\kappa)^3$ and $m/\kappa$ through the dependence of the Bohr radius on the effective mass. In addition, the Mott transition critical density is independent of $K$ or $n_i$ whereas the percolation transition density goes as $\sim n_i^\frac23$ in \equref{eq:perc}. Also, both Mott and percolation critical densities are manifestly independent of the valley or spin degeneracy.

To summarize, we have analyzed the various high density and low density limits of the Anderson localization critical density using the quantum and transport IRM criterion. Experimentally, bulk semiconductors are always at a finite compensation where the carrier density is lower than the impurity density ($n<n_i$). To make a realistic connection with experiments, in the next section, we provide analytical results for the critical density as a function of the compensation $K$ and analyze the high compensation and low compensation behaviors of the critical density.

\section{Critical density as a function of Compensation}
\label{sec:analytic}
Having described the behavior of the critical density as a function of impurity density in various regimes of density, we now provide analytical results for the critical density as a function of compensation $K$, and compare the analytical results with the exact numerically calculated $n_c$. Finally, we compare the critical densities from various mechanisms --- Mott, Anderson and percolation --- to predict the regions of compensation where we expect the MIT to be primarily of the Anderson localization type. We emphasize that the Mott transition critical density is completely independent of compensation since it depends only on the Bohr radius.

We first calculate $n_c$ as a function of $K$ using the quantum IRM, because the analytical results are tractable as compared with transport IRM, since $\tau_t$ is log-singular in $q_s$ for small $q_s$. From the quantum IRM criterion $\tau_q\epsilon_F=\hbar$, we find $n_i$ as a function of $n_c$ to be 
\beq
n_i = \frac{\hbar k_F^2/2m}{\frac{m}{8\pi\hbar^3k_F^3}\left(\frac{4\pi e^2}{\kappa^2}\right)^2\int_0^1 dx \frac{x}{(x^2+q_s^2F(x))^2}} = f_{q}(n_c).
\eeq
We first focus on the small $K$ limit. The zero compensation $K=0$ critical density is defined by the fixed point of $f$, $f_{q}(n_c)=n_i=n_c$, which we call $n^*$. Slightly away from $n^*$, we define $n_i=n^*+\delta n_i$ and $n_c=n^*+\delta n_c$ satisfying $n_i=f_q(n_c)$. Series expanding $f_q(n_c)$ to first order in $\delta n_c$ gives 
\beq
\delta n_i \approx \partial_n f_{q}(n)|_{n=n^*} \delta n_c = D_{q}\delta n_c,
\eeq
where we define $D_q=\partial_n f_{q}(n)_{n=n^*}$, which is given by 
\beq
D_{q}= \frac53 - \frac23 q_s^2\frac{1}{\int_0^1 dx \frac{x}{(x^2+q_s^2F(x))^2}}\int_0^1 dx \frac{xF(x)}{(x^2+q_s^2F(x))^3},
\eeq
where the expression for $D_q$ is evaluated at $n=n^*$. Using $\delta n_i = D_{q}\delta n_c$, and $n_i=n_c/(1-K)$, we write $n_c$ in terms of $K$ to find 
\beq
n_c =\frac{n^*(1-K)}{1+\frac{D_{q}}{1-D_{q}}K}\approx n^*(1-\frac{1}{1-D_{q}}K)\approx n^*e^{-\frac{1}{1-D_{q}}K}.\label{eq:kanalsmallquantum}
\eeq
In the above equation, we presented $n_c$ as a linear function of $K$, and as an exponential function of $K$. Though both of the expressions are consistent with each other in the limit of $K\to 0$, we find that the exponential function agrees better with the numerical calculation of $n_c$ for larger ranges of $K$. 

Next, we consider the limit of $K\sim 1$ ($q_s\ll 1$). In this regime, we approximate the integral inside the expression for $\tau_q$ by setting $F(x)=1$. The integral reduces to 
\beq
\int_0^1 dx\frac{x}{(x^2+q_a^2F(x))^2} \approx \frac12\frac{1}{q_s^2(q_s^2+1)}.
\eeq
Thus, the expression for $n_i$ becomes 
\beq
n_i =  \frac{\hbar^2k_F^2/2m}{\frac{m}{8\pi\hbar^3k_F^3}\left(\frac{4\pi e^2}{\kappa^2}\right)^2\int_0^1 dx \frac{x}{(x^2+q_s^2F(x))^2}} = \eta n_c^{\frac43}(\zeta n_c^{-\frac13}+1),
\eeq
where we define
\beq
\eta = \frac{4\hbar^2ge^2}{m\kappa \left(\frac{4\pi e^2}{\kappa}\right)^2}\left(\frac{6\pi^2}{g}\right)^\frac43, \phantom{11}\zeta = \frac{gme^2}{2\pi\kappa\hbar^2}\left(\frac{g}{6\pi^2}\right)^\frac13.
\eeq
Using $n_i=n_c/(1-K)$, we simplify the expression for $n_i(n_c)$ to get
\beq
n_c = \left(\frac{1}{\eta(1-K)}-\zeta\right)^3 \approx \frac{1}{\eta^3(1-K)^3}.\label{eq:kanallargequantum}
\eeq
\begin{figure}[tb]
    \centering
    \includegraphics[width=\columnwidth]{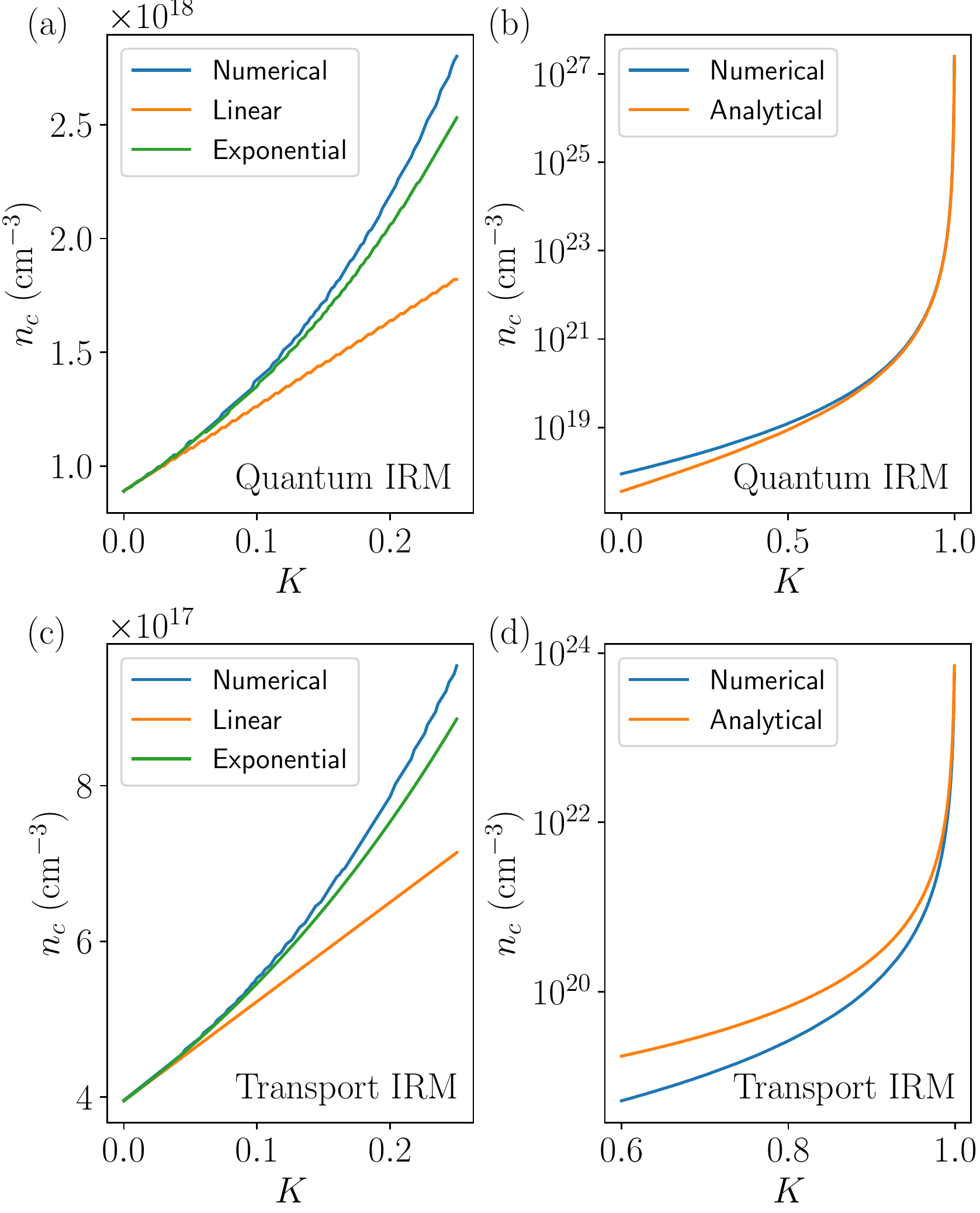}
    \caption{(a) Comparison of analytic formulas from \equref{eq:kanalsmallquantum} with numerically calculated $n_c$ as a function of compensation $K$ for quantum IRM. We see that for small $K$, the exponential formula agrees well with the numerically calculated result. (b) Comparison of large $K$ analytic formula \equref{eq:kanallargequantum} for $n_c(K)$ with the numerical result for quantum IRM. We see that the analytic formula agrees well with the numerical result all the way down to $K=0$. (c) Same as (a), for transport IRM with the analytic formula from \equref{eq:kanalsmalltransport}, we see that the exponential formula is in good agreement with the numerical result. (d) Same as (b), with transport IRM using the analytic formula from \equref{eq:kanallargetransport}. While the analytic result matches the numerical result for large $K$, the agreement is not very good for small $K$ since we ignored the $\ln\ln n_c$ terms in the analytical formula.}
    \label{fig:kanal}
\end{figure}

We repeat the same analysis for the transport IRM case to derive the large $K$ and small $K$ analytic behavior. The corresponding definition of $f_t(n_c)$ is given by,
\beq
n_i = \frac{\hbar k_F^2/2m}{\frac{m}{8\pi\hbar^3k_F^3}\left(\frac{4\pi e^2}{\kappa^2}\right)^2\int_0^1 dx \frac{2x^3}{(x^2+q_s^2F(x))^2}} = f_{t}(n_c).
\eeq
The analysis of the low $K$ behavior proceeds similarly to the quantum IRM case, which gives us 
\beq
n_c =\frac{n^*(1-K)}{1+\frac{D_{t}}{1-D_{t}}K}\approx n^*(1-\frac{1}{1-D_{t}}K)\approx n^*e^{-\frac{1}{1-D_{t}}K},\label{eq:kanalsmalltransport}
\eeq
where we define
\beq
D_{t}= \frac53 - \frac23 q_s^2\frac{1}{\int_0^1 dx \frac{x^3}{(x^2+q_s^2F(x))^2}}\int_0^1 dx \frac{x^3F(x)}{(x^2+q_s^2F(x))^3}.
\eeq
The large $K$ behavior, on the other hand, is not as straightforward as in the quantum IRM case. This is because in the $q_s\ll 1$ limit, the integral in the expression defining $\tau_t$ reduces to 
\beq
\int_0^1 dx\frac{x^3}{(x^2+q_s^2F(x))^2}\approx -\ln q_s,
\eeq
which contains a $\ln q_s$ singularity. The corresponding expression for $n_i$ becomes 
\beq
n_i = \frac{2\pi\hbar^4\left(\frac{6\pi^2}{g}\right)^\frac53}{m^2\left(\frac{4\pi e^2}{\kappa}\right)^2}\frac{n_c^\frac53}{-\delta+\frac16 \ln n_c},
\eeq
and using $n_i=n_c/(1-K)$, the expression simplifies to 
\beq
1-K = \gamma \frac{\frac16\ln n_c -\delta}{n_c^\frac23},
\eeq
where we define
\beq
\gamma = \frac{m^2\left(\frac{4\pi e^2}{\kappa}\right)^2}{2\pi\hbar^4\left(\frac{6\pi^2}{g}\right)^\frac53}.
\eeq
The expression for $n_c$ cannot be analytically inverted due to the $\ln n_c$ term, however, we can perform an iterative approximate inversion as follows
\beq
n_c = \left(\gamma\frac{\frac16\ln n_c - \delta}{1-K}\right)^\frac32 \approx \left(\gamma\frac{\frac14\ln\left(\frac{\gamma}{1-K}\right) - \delta}{1-K}\right)^\frac32.\label{eq:kanallargetransport}
\eeq
In performing the approximate analytical inversion to find $n_c$ as a function of $K$, we ignored terms of the order $\ln\ln n_c$.

\begin{figure}[tb]
    \centering
    \includegraphics[width=\columnwidth]{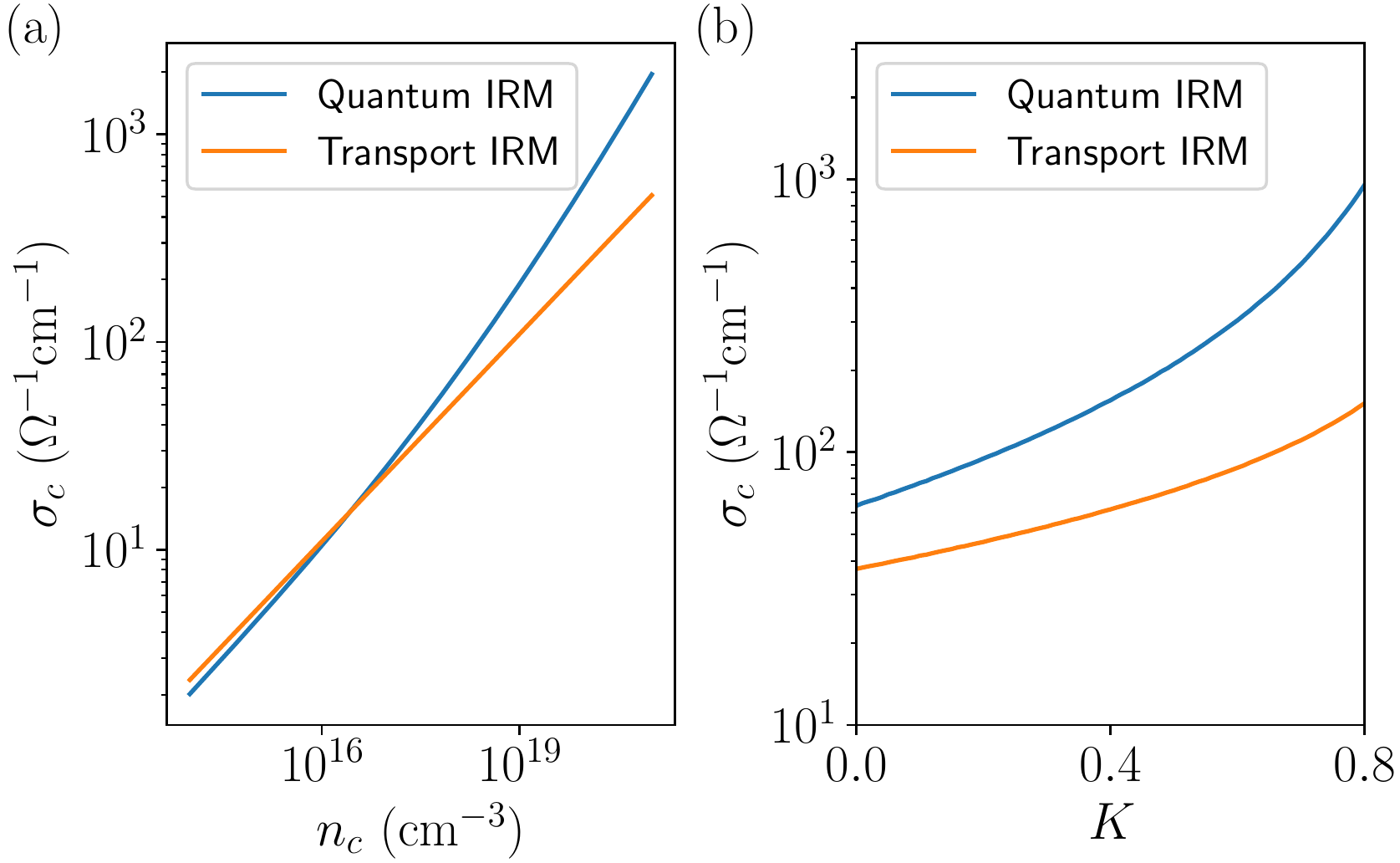}
    \caption{Conductivity $\sigma_c$ at the Anderson localization point as a function of (a) $n_c$ and (b) compensation $K$. (a) We see that as a function of $n_c$, $\sigma_c$ calculated from transport IRM follows $\sigma_{c,t}\sim n_c^\frac13$ for all $n_c$. On the other hand, while $\sigma_c$ calculated from quantum IRM follows $\sigma_{c,q}\sim n_c^\frac13$ for small $n_c$, it increases at a faster rate for large $n_c$ and follows $\sigma_{c,q}\sim n_c^\frac23$. (b) As a function of compensation $K$, we are never in the small $n_c$ regime, and $\sigma_{c,q}>\sigma_{c,t}$ is always true.  }
    \label{fig:sigmac}
\end{figure}

With the analytic dependence of $n_c$ as a function of $K$, we proceed to provide a comparison of the analytic results with the numerically calculated exact values of $n_c$. In \figref{fig:kanal}(a), we show the comparison of the small $K$ numerically calculated $n_c$ from quantum IRM as a function of $K$, with the analytic formula in \equref{eq:kanalsmallquantum}. We see that the exponential analytic formula agrees better with the numerical result for a larger range of compensation as compared to the linearized function. In \figref{fig:kanal}(b), we show the comparison of the corresponding large $K$ analytic formula \equref{eq:kanallargequantum} with the corresponding numerical result. We see that the agreement of the analytic result is very good with the numerical result, all the way up to small $K$. In \figref{fig:kanal}(c) and (d), we show the comparison of the small $K$ and large $K$ numerical result using transport IRM, with the corresponding analytic results based on \equref{eq:kanalsmalltransport} and \equref{eq:kanallargetransport} respectively. In the regime of low compensation, we find that the agreement of the exponential formula is almost exact up to compensation of $K\sim 0.25$. The large $K$ analytical formula, on the other hand, overestimates $n_c$ by a factor of about $\sim \sqrt{10}$ as compared to the numerical result. This disagreement is caused by ignoring the $\ln\ln n$ corrections in \equref{eq:kanallargetransport}, however, the disagreement vanishes if we look at the close vicinity of $K\sim 1$.

\begin{figure}[tb]
    \centering
    \includegraphics[width=\columnwidth]{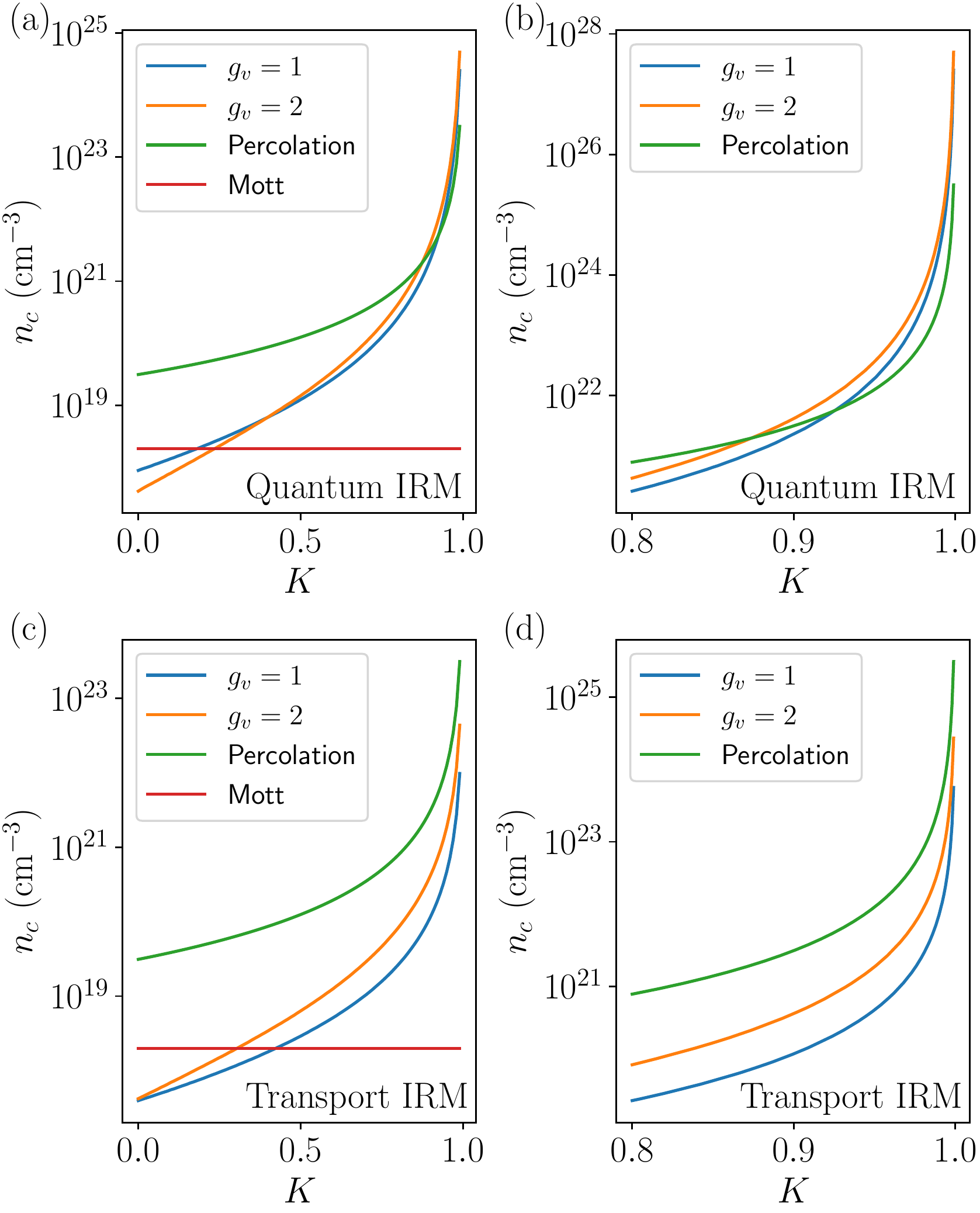}
    \caption{Comparison of $n_c$ (for valley degneracy $g_v=1$ and $g_v=2$), Mott density $n_M$ and percolation transition density $n_p$ (a) using quantum IRM, for all $K$, (b) using quantum IRM, zoomed into large $K$ region. (c) and (d) are the same as (a) and (b) respectively, with the corresponding $n_c$ calculated using transport IRM. For very small $K$, the Mott transition is the leading mechanism for both quantum and transport IRM. Near $K\sim 0.2$, the IRM mechanism becomes relevant for the MIT. For large $K>0.8$, there is close competition between the percolation transition and IRM mechanism. At very large $K$, as we see in (b), the quantum IRM mechanism wins over the percolation transition as the leading mechanism, however (as we see in (d)) the percolation transition closely wins over the transport IRM mechanism. }
    \label{fig:nccomparison}
\end{figure}
We also calculate the minimum critical conductivity, $\sigma_c=\sigma(n=n_c)$, at the Anderson localization point, defined by \cite{mott1967electrons}
\beq
\sigma_c= n_c e^2\tau_t(n_c)/m.
\eeq
Note that conductivity is always defined using the transport scattering time. Using the transport IRM, at the Anderson localization point, the transport scattering time is given by $\tau_t= \hbar/\epsilon_F$, and the critical conductivity becomes 
\beq
\sigma_{c,t} = n_c e^2\hbar/(m\epsilon_F)  =\frac{e^2}{\hbar} \frac{\pi}{\left(6\pi^2/g\right)^\frac23}n_c^\frac13.
\eeq
The corresponding formula for $\sigma_{c,q}$ using the quantum IRM is not straightforward, because the critical density $n_c$ should be calculated using the quantum IRM, and then the calculated value of $n_c$ should be used to find the transport scattering time $\tau_t$.  In \figref{fig:sigmac}(a) and (b), we show the critical conductivity for both the quantum and transport IRM as a function of $n_c$ and compensation $K$ respectively. While for low density ($q_s\gg 1$) $\sigma_{c,q}$ and $\sigma_{c,t}$ agree, they differ for large density ($q_s\ll 1$) where $\sigma_{c,q}\sim \frac{n_c^\frac23}{\ln n_c}$. 

While the above calculation for $\sigma_c$ is done at the MIT critical density $n_c$, we additionally calculate the scaling form of the conductivity for a generic carrier density $n$ and impurity density $n_i$, on the metallic side of the MIT. In the low density limit (which is relevant close to the MIT), the transport scattering time is asymptotically given by 
\begin{align}
    \tau_t\sim k_F^3 q_s^4/n_i \sim n^{1-\frac{4}{6}}/n_i\sim n^\frac13/n_i.
\end{align}
Thus, the conductivity scales as
\begin{align}
    \sigma(n) = n e^2\tau_t/m \sim n^\frac43/n_i.
\end{align}
We emphasize that the above expression is simply a power law exponent for the growth of the conductivity as a function of carrier density, and not a critical exponent (if it exists) for the MIT. On the other end of large carrier density, the transport time is given by $\tau_t\sim -k_F^3/\ln q_s n_i \sim n/n_i\ln n $, and the corresponding conductivity scales as $\sigma(n) \sim n^2/n_i\ln n $. These power laws (with the conductivity exponent varying from $4/3$ at low density to $2$ at high density) that we derive is calculated using Boltzmann transport, which is valid only on the metallic side of the metal-insulator transition. They are valid in the regime $n>n_c$, and are in agreement with some experiments\cite{v2011electron}. 

Finally, in \figref{fig:nccomparison}, we show the comparison of $n_c$ (calculated using the IRM criterion), $n_M$ (calculated using the Mott criterion), and $n_p$ (calculated using the percolation localization). In terms of compensation, the expression for $n_p$ is given by 
\beq 
n_p = \left(\frac{0.7}{a_B}\right)^3\frac{1}{(1-K)^2}.
\eeq
Though we have shown $n_p$ for all ranges of compensation, the percolation mechanism for MIT is valid only at high compensations. \figref{fig:nccomparison}(a) and (b) shows the comparison with $n_c$ calculated based on quantum IRM, with (b) showing the comparison between $n_c$ and $n_p$ at large $K$. Since $n_c\sim \frac{1}{(1-K)^3}$, as expected the quantum IRM density becomes larger than the percolation density. At small $K$, the Mott mechanism dominates over the quantum IRM till $K\sim 0.2$, after which the IRM criterion becomes the main mechanism for the MIT. \figref{fig:nccomparison}(c) and (d) show the same results as \figref{fig:nccomparison}(a) and (b), but with $n_c$ calculated using the transport IRM instead of the quantum IRM. At small $K$, the Mott transition remains dominant till a larger value of compensation (as compared to the quantum IRM). Increasing the valley degeneracy makes the Transport IRM criterion the dominant mechanism for lower compensations. For large $K\sim 1$, the percolation transition prevails over the transport IRM $n_c$, because for the transport IRM, $n_c\sim \frac{1}{(1-K)^\frac32}$ near $K\sim 1$, while the percolation transition density goes as $n_p\sim \frac{1}{(1-K)^2}$. As a result, near $K\sim 1$, $n_p\gg n_c$. We mention, however, that the percolation transition is based on semiclassical arguments involving conducting paths through the inhomogeneous puddle disorder landscape of 'lakes and mountains', and it is likely that at $T=0$, once quantum tunneling through the 'mountains' becomes dominant, the percolation transition crosses over to the Anderson localization transition since both are disorder-dominated transition from a conductor to an insulator. Further discussion of percolation is well beyond the scope of the current work where our focus is entirely on the crossover conductor-to-insulator transition density arising from the IRM coherence criterion.

\section{Anderson Localization for Metals}

\begin{figure}[tb]
    \centering
    \includegraphics[width=\columnwidth]{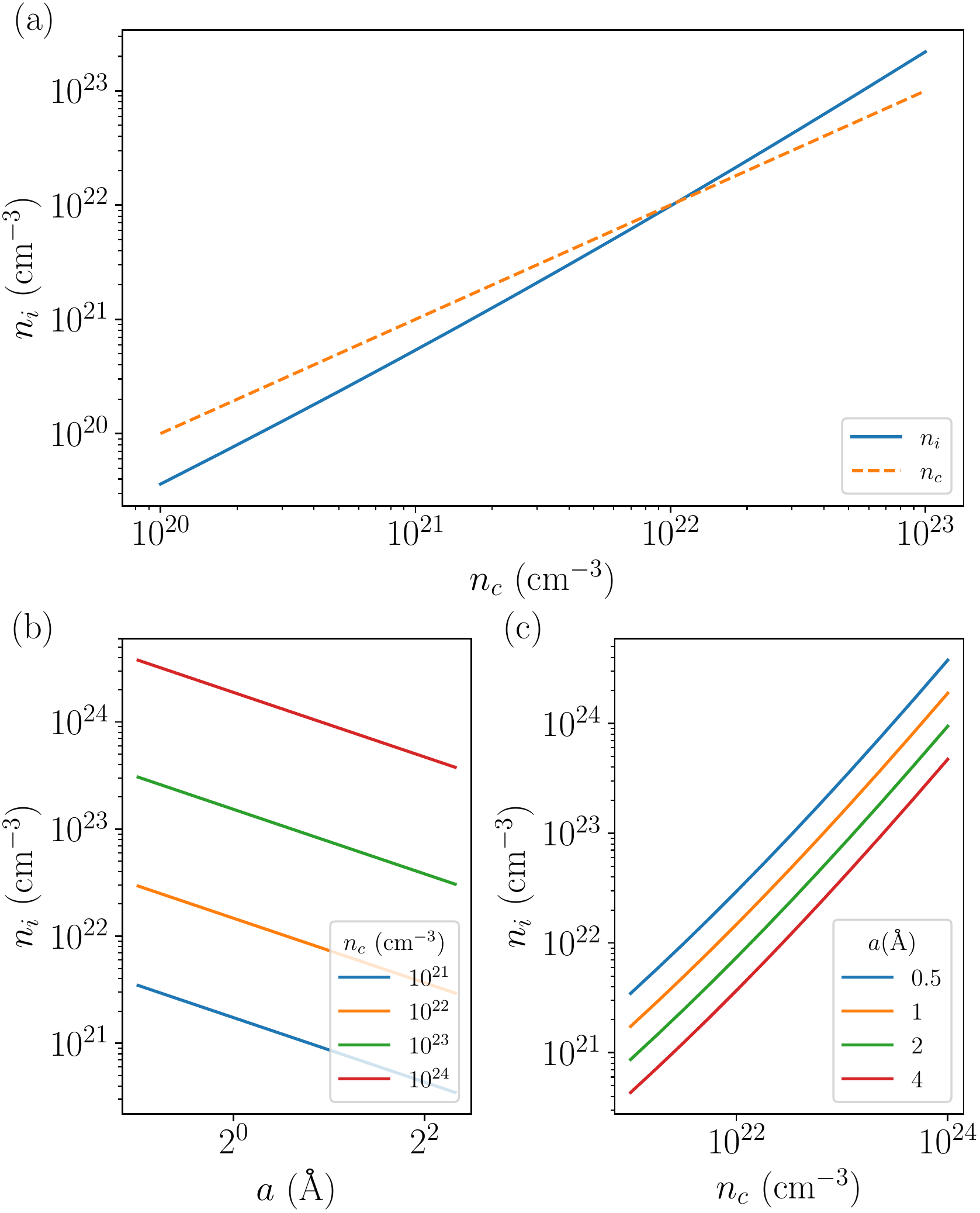}
    \caption{(a) Impurity density $n_i$ as a function of critical carrier density $n_c$ from the IRM criterion $l_{MFP}=k_F^{-1}$ (b) $n_i$ as a function of lattice constant $a$ for various values of $n_c$, using the IRM criterion for metals $l_{MFP}=a$ (c) $n_i$ as a function of $n_c$ for various values of lattice constant $a$, using the IRM criterion for metals.}
    \label{fig:fig7}
\end{figure}

Before we conclude, we provide a brief discussion on the existence of disordered induced Anderson localization in metals. Regular metals have a large carrier density, with the Fermi level inside the conduction band. The Fermi momentum (wavelength) of metals is large (small). An important characteristic of metals is that they completely screen the charged impurity potential, strongly suppressing the effects of random charged disorder. Effectively, the dielectric screening constant becomes 
\begin{align}
    \epsilon(q) \approx \left(q_{TF}/q\right)^2 F(q/2k_F),
\end{align} 
and the resulting screened impurity potential given by
\begin{align}
    u_{\bq} = \frac{4\pi e^2}{q_{TF}^2F(q/2k_F)}\approx \frac{4\pi e^2}{q_{TF}^2},
\end{align} 
where we work in the small $\bq$ limit, and additionally set $\kappa=1$ for metals. The resulting impurity density required for localizing the electrons is given by the same as Eq.~\ref{eq:ncnilowden}, with 
\begin{align}
    n_i \sim n_c/4.
\end{align}
\figref{fig:fig7}(a) shows the numerically exact solution of the IRM criterion (Eq.~\ref{eq:irm}) for the range of densities relevant to metals. We note that the required impurity density to localize the electrons is of the same order of magnitude as that of the carrier density, and is larger than the carrier density for $n_c\gtrsim 10^{22}{\rm cm^{-3}}$. This is an unphysically high impurity density (approximately one impurity per unit cell), making it essentially impossible to Anderson localize regular metals, and indeed the Anderson localization has never been observed in regular high-density metals.

Due to the large carrier density, the Fermi wavelength in metals is often very small, with $k_F^{-1}\sim 10^{-8}{\rm cm}$. In such situations, the Fermi wavelength can end up being smaller than the lattice constant $a$ of the crystalline metal. For coherent transport, however, the electron wavefunction needs to be coherent not just over the length scale of the Fermi wavelength, but also over the length scale of the crystalline structure. As a result, for regular metals, the IRM criterion is often restated as \cite{hussey2004universality,ioffe1960non,joffe1956heat,gurvitch1981ioffe}
\begin{align}
    l_{MFP} &= a \text{\phantom{=====}} a\ge k_F^{-1},\\
    l_{MFP} &= k_F^{-1} \text{\phantom{===}} a<k_F^{-1}.
\end{align}
With the IRM criterion $l_{MFP}=a$, we plot the resulting impurity density as a function of lattice constant $a$ for various values of $n_c$ in Fig.~\ref{fig:fig7}(b). We show the required impurity density as a function of critical density for various values of lattice constants in Fig.~\ref{fig:fig7}(c). The required impurity density is very large ($>10^{23} {\rm cm^{-3}}$), and unphysical when $n\gtrsim 10^{23}{\rm cm^{-3}}$. In metals and semi-metals that have relatively low carrier density ($n\sim 10^{20-21} {\rm cm^{-3}}$), it is indeed possible to localize the electrons with a very large amount of disorder. Such a situation has been observed in iron compounds\cite{ying2016anderson,williams2009metal}, cuprates\cite{mikhajlov1989anderson,acharya2018metal,orgiani2007direct}, and graphene\cite{eda2009insulator} among others\cite{imada1998metal}. By contrast, regular high density metals cannot be localized by disorder since the necessary amount of disorder is unphysical.

\section{Conclusion}
\label{sec:conclusion}

\begin{table}[tb]
\begin{center}
\caption{Summary of our analytical results results at the low density $(q_s\gg 1)$ and high density $(q_s\ll 1)$ limits}
\label{tab:resultssummary}
\begin{ruledtabular}\begin{tabular}{ccc} 
IRM & Quantum IRM & Transport IRM\\\hline
$n_c (q_s\gg 1)$& $\sim  \frac{8\pi C_q}{3g} n_i$ & $\sim\frac{8\pi C_t}{3g} n_i$\\
$n_c (q_s\ll 1)$& $\sim\lt(g^\frac13 \frac{m}{\kappa} n_i\rt)^\frac34$& $\sim g\lt(\frac{m^2}{\kappa^2} n_i\rt)^\frac35$ \\[2ex]\hline
$8\pi C_{t,q}/3$  &$\int_0^1 dx~ x/F(x)^2\sim 7.3$ & $\int_0^1 dx~ 2 x^3/F(x)^2\sim 8.8$\\[2ex]\hline
$n_c (K\ll 1)$ &$\sim n^* \exp\lt(-\frac{1}{1-D_q}K\rt)$ & $\sim n^* \exp\lt(-\frac{1}{1-D_t}K\rt)$\\
$n_c(K\sim 1)$ &$\sim {1}/{\eta^3(1-K)^3}$ &$\sim {\gamma^\frac32}/{(1-K)^\frac32}\ $\\[2ex]\hline
$\sigma_c (q_s\gg 1)$ & $\sim g^\frac23  n_c^\frac13$&$\sim g^\frac23  n_c^\frac13$ \\
$\sigma_c (q_s\ll 1)$ & $\sim   \frac{\kappa}{m g^{2/3}} n_c^\frac23/\ln n_c$ &$\sim g^\frac23  n_c^\frac13$\\[2ex]
 \end{tabular}
\end{ruledtabular}
\end{center}
\end{table}

We develop a theory for the doping induced metal-insulator transition in bulk semiconductors assuming the transition to be a quantum Anderson localization, as defined by the IRM criterion for the loss of coherent transport, due to carrier scattering by screened Coulomb disorder associated with spatially randomly localized quenched ionized dopant atoms. Our theory provides specific analytical predictions for the dependence of the critical density $n_c$ on the random charged impurity density with $n_c\sim n_i$ and $n_c\sim n_i^{\frac35}$ in the low-($q_{s}\gg 1$) and high-($q_{s}\ll 1$) density limit, respectively, using the transport IRM. A key feature of our theory is a specific prediction on how the critical density varies with compensation $K:n_c\sim\frac{1}{(1-K)^\frac32}$ using the transport IRM. A summary of our analytic results is shown in Table~\ref{tab:resultssummary}. We note that $n_c$ diverges as $K$ approaches unity since the metallic phase is totally suppressed for complete compensation, and the system is thus always a trivial insulator with no conducting carriers. Our predicted $K$ dependence of the MIT has been observed experimentally, establishing the relevance of the Anderson localization scenario to the MIT phenomenon on semiconductors\cite{thomanschefsky1992metal}. We mention that the competing strict Mott transition scenario does not predict any compensation dependence of the critical density which is universal in the Mott transition scenario independent of $K$. We believe that the Anderson localization picture applies increasingly more accurately with increasing compensation whereas the Mott transition picture perhaps applies better to strictly uncompensated systems \cite{shklovskiiprivatecomm,ghazali1978density,newman1983metal,v2011electron}. There is no independent way of ensuring a sample to be uncompensated since the unintentional compensation due to the unknown presence of both donors and acceptors is more likely generic. Since the randomness is enhanced for higher compensation, it is likely that the doping induced MIT crosses over from being more Mott-like for uncompensated semiconductors to being more Anderson-like for compensated semiconductors.

Our theory uses Boltzmann transport to calculate the transport scattering time and conductivity. However, Boltzmann transport is valid only in the metallic regime, and the insulating transition never shows up within this framework. Instead, we approach the MIT from the metallic side, where Boltzmann transport is valid, and operationally define the MIT using the IRM criterion. Near the IRM point, Boltzmann transport is not expected to be quantitatively valid. On the other hand, the IRM criterion allows us to derive a large collection of analytical results, trends and power-laws, which are expected to be qualitatively correct. Our theory is by no means the complete theory of Anderson localization, but it nonetheless provides a lot of insight about the behavior and properties of the metal-insulator transition in semiconductors. We emphasize that our MIT is synonymous with IRM, and we uncritically assume that the IRM criterion gives the condition for MIT.  Our theory would not describe any MIT which happens far away from the IRM condition.

The theory is at $T=0$. At finite, but low temperatures, ignoring the exponentially small thermal carrier excitation, the main effects would be a thermal weakening of screening and a thermal smearing of the Fermi surface, which oppose each other in affecting the mean free path (and the effective scattering rate). In three dimensional semiconductors, the thermal averaging effect wins out over reduced screening, producing an $O(T^2)$ increase in the effective mean free path, consequently leading to a decreasing $n_c$ with increasing temperature (keeping all other parameters fixed). Of course, with further increase in temperature, thermal occupancy of ionized carriers complicates the picture as the system becomes `trivially' metallic except the conductivity is thermally activated akin to insulators. We note that for the Mott transition, however, the only finite temperature effect would be a suppressed screening, leading consequently to an increasing critical MIT density with increasing temperature, again the increase is a weak $O(T^2)$ effect. The difference between the thermal effects on the critical density between the Anderson localization and the Mott transition can, in principle, be used to distinguish which mechanism may be operational in a particular observation of the doping induced MIT. 

We note that our theory can only obtain the critical density for the MIT, and cannot predict the critical exponent since our theory is not a critical (or scaling) theory, but is a theory for the doping induced crossover between a metal and an insulator. The IRM criterion describes a crossover and not a transition, and is thus well-suited to predict the transition or the crossover density, but not any critical behavior of the conductivity near the transition point. Our theory, being based on the Boltzmann transport theory, predicts a finite resistivity at the crossover point defined by $n_c$, and the IRM criterion asserts that the system is a metal (with increasing resistivity with increasing temperature) for $n>n_c$ and an insulator (with decreasing resistivity with increasing temperature) for $n<n_c$\cite{mott1972conduction}. There is no obvious critical scaling behavior of the conductivity in our theory since we use the Boltzmann transport formalism to calculate the mean free path. By contrast, critical scaling theories cannot predict the value of the critical density, which is the focus of the current work. It is possible to calculate the critical point based on numerical scaling techniques for simple models\cite{markos2006numerical}, however, taking into account realistic Coulomb disorder potential remains computationally challenging. Also, calculating the critical density itself is well beyond the scope of numerical scaling localization calculations, and the calculation of $n_c$ as a function of system parameters is the main point addressed in our work. Experimentally, $n_c$ is much easier to determine than the scaling exponents\cite{hirsch1988critical,stupp1993possible}, but our theory is certainly limited, because of the use of the IRM criterion in treating the MIT as a crossover at $n_c$. The existence of a minimal metallic conductivity at the MIT remains controversial. Theories based on the scaling theory of localization find that the MIT should be continuous, without a minimum value of the conductivity\cite{mcmillan1981scaling}, which has been verified by some experiments\cite{dodson1981metal,zabrodskii1984low}. However, a sharp MIT has also been observed in semiconductor systems\cite{rosenbaum1980sharp}, including phase change materials\cite{siegrist2011disorder,bragaglia2016metal,oka2021electron}. The challenge in distinguishing a sharp MIT with a minimal conductivity from a continuous MIT is in extrapolating the finite temperature conductivity to find the corresponding $T=0$ value, since there is no clear-cut and universal way to perform the extrapolation. In fact, there are double extrapolations involved in the study of the MIT as a continuous transition: One must extrapolate $\sigma(n, T)$ to $T=0$ for each $n$, and then extrapolate the density to $n=n_c$ in order to ascertain where $\sigma(n=n_c; T=0)$ vanishes. such extrapolations are inherently error-prone, particularly since $\sigma(n)$ for $n\sim n_c$ is extremely small. Recently, it has also been noted that different methods of extrapolation can lead to differing conclusions with regard to the nature of the transition\cite{mobius2019metal}.

The critical density $n_c$ in our IRM-based theory has a physical explanation, which requires no extrapolation and is thus experimentally more appealing. The conductivity $\sigma(n, T)$ does not vanish at the critical density $n_c$ we obtain, but $d\sigma/dT$ changes its sign at $n=n_c$ from being negative for $n>n_c$ to being positive for $n<n_c$ at low enough temperatures. The basic idea is that a metal (insulator) has a conductivity decreasing (increasing) with temperature since at $T=0$ a metal (insulator) has finite (zero) conductivity. In fact, the definition of a metal versus an insulator in terms of the temperature coefficient of the conductivity is well-established, and operationally effective since it requires neither an extrapolation to $T=0$ nor an extrapolation in density (in extracting the critical density). Accepting that the `critical' density we calculate is the crossover density where the derivative of the conductivity changes its sign side-steps the controversial issue of the existence or not of a minimum metallic conductivity since $n_c$ now is a crossover density defining a crossover from an effective insulator for $n<n_c$ (with $d\sigma/dT > 0$) to an effective metal for $n>n_c$ (with $d\sigma/dT <0$). The possibility that the actual critical density for the MIT is lower than our calculated $n_c$ cannot be ruled out, but our calculated $n_c$ is certainly close to any putative transition density (if the MIT is indeed a continuous transition with $\sigma(n)$ vanishing at the critical density) since $\sigma(n)$ varies fast near $n\sim n_c$. If $n_c$ is indeed the crossover density where $d\sigma/dT$ changes the sign, there is no conceptual problem with $\sigma(n=n_c)$ being finite since it is only the conductivity at $n=n_c$ which, if the transition is indeed continuous, slightly on the metallic side of the transition.  

Since our theory is not a critical theory for the localization transition, the concept of a critical exponent, $\nu$, where $\sigma(n) \sim (n-n_c)^\nu$ close to the localization transition, does not apply. The value of the localization exponent and whether such an exponent even exists (i.e. whether the transition is continuous or not) have been much discussed in the literature, but we have nothing to add to this debate since ours is a theory for the crossover associated with the loss of metallic coherence and not a theory for the scaling localization itself. We can, however, calculate an exponent `$\nu$' defined by $\sigma(n) \sim n^\nu$ where our $\nu$ simply defines the power law growth of the 3D conductivity with carrier density. It is straightforward to show that for Coulomb disorder at low density (where $q_s\gg 1$), the conductivity exponent is $1.33$\cite{sarma2013universal}. Thus, we predict that $\nu=1.33$, which is in agreement with some experiments\cite{v2011electron}, but not with others\cite{rosenbaum1980sharp}. We emphasize that our calculated exponent is not a critical scaling exponent, but a power law exponent for the density dependence of the conductivity. It may be interesting to note that our low-density conductivity exponent ($\nu=1.33$) for Coulomb disorder is actually very close to the theoretical critical exponent for the 3D Anderson localization in correlated disorder as calculated by direct numerical diagonalization on a tight binding lattice, finding $\nu \sim 1.3-1.5$\cite{croy2012role}. Whether this is a pure coincidence with no significance or not is unknown, but we believe that the conductivity $\sigma(n)$ should scale as $\sim n^{1.33}$ in the low-temperature metallic doped semiconductors for $n\gg n_c$. This prediction should be experimentally tested. Note that our theory predicts a density dependence of the crossover exponent $\nu(n)$, with $\nu$ increasing from $1.33$ for $q_s\gg 1$ (low density) to $\nu=2$ for $q_s\ll 1$ (high density)\cite{sarma2013universal}. 

Our theory uses many approximations, in addition to assuming the MIT to be strictly a localization phenomenon. We use RPA for screening the bare Coulomb disorder, which should be adequate, but by no means exact. We use the Boltzmann transport theory to calculate the mean free path, which is reasonable when the mean free path is much larger than $1/k_F$, but fails in an unknown way close to the transition point where $l_{MFP}=1/k_F$. Since the Boltzmann theory remains well defined for all values of $l_{MFP}$ as long as $n>n_c$, the approximation of using it all the way to the IRM point of $l_{MFP} \cdot k_F=1$ introduces quantitative, but not qualitative, errors. Going beyond our approximation including higher order scattering diagrams would not improve the situation much since the scattering regime close to the transition is most likely nonperturbative, necessitating a fully numerical localization calculation, which is beyond the scope of the current work where our interest in understanding general theoretical trends, and in particular, the behavior of the MIT density $n_c$ on system parameters. In any case, such a fully numerical exact localization calculation taking into account long-range correlated Coulomb disorder is computationally challenging and has not been attempted in the literature. The goal of the current work is analytical understanding of the critical density on Coulomb disorder rather than numerical precision. 

A relevant question is how one can ascertain whether a particular doping induced MIT in an experiment is more Mott-like or more Anderson-like since the two theories emphasize complementary physical mechanisms for the MIT (interaction for Mott and disorder for Anderson). In general, as already stated above, we believe that with increasing compensation, disorder is enhanced, and therefore, the MIT becomes more like an Anderson localization at higher compensations\cite{thomas1982evidence}. In reality, of course, both interaction and disorder are important, and a sharp distinction between Mott and Anderson is not meaningful with the MIT being an Anderson-Mott transition\cite{rosenbaum1983metal,byczuk2005mott}. One simple rule to discern the two mechanisms could be that whichever has a higher critical density in a particular situation should win out since at low densities the system is insulating for both mechanisms\cite{siegrist2011disorder,bragaglia2016metal,oka2021electron}. This simple rule could in principle distinguish if a particular observed MIT is more Mott-like or more Anderson-like. Also, the Mott critical density is universal for each semiconductor as it depends only on the Bohr radius whereas the Anderson critical density varies with all system parameters, particularly the compensation. Any compensation dependence of $n_c$ argues in favor of the Anderson transition\cite{thomanschefsky1992metal}.

In principle, an Anderson localization type MIT can also be defined for metals and semi-metals. However, metals screen charged impurities completely, and therefore, the effects of charged impurity disorder is strongly suppressed. Moreover, in most metals (with a typical carrier density of $n\sim 10^{23} {\rm cm^{-3}}$), the Fermi momentum is of the order $k_F\sim 10^{8}{\rm cm^{-1}}$.  If it were possible to localize the electrons in a metal, then the corresponding mean free path required by the IRM criterion is $l_{MFP}\sim k_F^{-1}\sim 10^{-8} {\rm cm}$, necessitating a disorder concentration of about $n_i\sim 2\cdot 10^{23}{\rm cm^{-3}}$ -- twice as much as the electron density, which is impossible to achieve! Even amorphous and liquid metals (which are in some sense maximally disordered due to lack of a crystalline structure) are still conducting. Thus, in practice, the amount of disorder required to localize electrons in metals is unphysically large, unless the metal has relatively a very low density.  Such a situation can occur, for example, in cuprates or semi-metals, which have low carrier density of $n\sim 10^{21}{\rm cm^{-3}}$, and can be localized with a large amount of disorder. Our theory provides a reasonable description for such low-density metals, and also explains why Anderson localization has never been observed in regular high-density metals.

We conclude by emphasizing that a full theory for the MIT must include both the interaction effects of inter-electron Coulomb correlations intrinsic to the Mott transition and the localization effects of the random Coulomb impurity disorder inherent to the Anderson transition. A theory including both to obtain the critical density remains intractable and is one of the long-lasting unsolved problems in physics. How to include both disorder and interaction in a theory on a nonperturbative equal footing when both are strong remains open, and our work shows that neglecting the localization effects uncritically is a mistake, certainly for highly compensated samples with strong Coulomb disorder effects.

\acknowledgements
This work is supported by the Laboratory for Physical Sciences.

\bibliography{refs.bib}

\end{document}